\documentclass{jpsj-suppl}
\usepackage{txfonts} 
\usepackage{graphicx,amsmath,wrapfig,epsfig}

\title{Internal structure of exotic hadrons by high-energy exclusive reactions}
\author{
H. \textsc{Kawamura}$^a$, S. \textsc{Kumano}$^{b,c}$, 
and T. \textsc{Sekihara}$^d$
}
\inst{
$^a$ 
Department of Mathematics, Juntendo University, Inzai, Chiba,
270-1695, Japan \\
$^b$KEK Theory Center, Institute of Particle and Nuclear Studies, KEK \\
\ \ 1-1, Ooho, Tsukuba, Ibaraki, 305-0801, Japan \\
$^c$J-PARC Branch, KEK Theory Center,
Institute of Particle and Nuclear Studies, KEK \\
\ \ and Theory Group, Particle and Nuclear Physics Division, 
J-PARC Center \\
\ \ 203-1, Shirakata, Tokai, Ibaraki, 319-1106, Japan \\
$^d$ Research Center for Nuclear Physics (RCNP), 
Osaka University, Ibaraki, Osaka, 567-0047, Japan

}

\recdate{September 12, 2014}

\abst{
We propose to use high-energy exclusive reactions for probing
internal structure of exotic hadron candidates.
First, the constituent counting rule of perturbative QCD 
can be used for finding internal configurations of
an exotic hadron candidate. It is because the number of constituents ($n$),
which participate in the exclusive reaction, is
found by the scaling behavior of the cross section
$d\sigma/dt \propto 1/s^{n-2}$ at large momentum transfer, 
where $s$ is the center-of-mass energy squared. 
As an example, we show that the internal structure of $\Lambda \, (1405)$
should be found, for example, by the reaction
$\pi^- + p \to K^0 + \Lambda (1405)$.
Second, the internal structure of exotic hadron candidates should be
investigated by hadron tomography with
generalized parton distributions (GPDs) and generalized
distribution amplitudes (GDAs) in exclusive reactions.
Exotic nature should be reflected
in the GPDs which contain two factors,
longitudinal parton distributions as indicated by
the constituent counting rule and transverse form factors as suggested 
by the hadron size. The GDAs should be investigated by
the two-photon process $\gamma^* \gamma \to h\bar h$,
for example $h=f_0$ or $a_0$, in electron-positron annihilation.
Since the GDAs contain information on a time-like form factor,
exotic nature should be found by studying
the $h\bar h$ invariant mass dependence of the cross section.
The internal structure of exotic hadron candidates should be clarified 
by the exclusive reactions at facilities such as J-PARC and KEKB.
}

\kword{hadron, exotic, exclusive, QCD, quark, gluon}

\begin{document}
\maketitle

\section{Introduction}

Existence of new hadrons has been investigated for a long time, 
and a few hundred hadrons have been discovered \cite{pdg-2014}. 
Almost all the hadrons are understood
by the internal configurations of $q\bar q$ and $qqq$ as proposed
in the original quark model. Exotic hadrons, such as tetraquark
($qq\bar q \bar q$), pentaquark ($qqqq\bar q$), and glueball ($gg$),
have been investigated since the quark-model proposal in 1964. 
For example, $f_0 (980)$ and $a_0 (980)$ are known as exotic hadron
candidates. In a native quark model, scalar mesons in the 1 GeV region
could be interpreted as 
$\sigma = f_0 (600) = (u\bar u+d\bar d)/\sqrt{2}$,
$f_0 (980) = s \bar s$, 
$a_0 (980) = u \bar d, \,  (u\bar u-d\bar d)/\sqrt{2}, \,
             d \bar u$.
Since the strange quark is heavier than up and down quarks,
these quark configurations suggest the mass hierarchy
$m (\sigma) \sim m (a_0) < m (f_0)$, which is in
contradiction to the experimental one
$m (\sigma) < m (a_0) \sim m (f_0)$, where $f_0 (980)$
is simply denoted as $f_0$. Furthermore, the strong-decay width
$f_0 \to 2\pi$ in the quark model largely overestimates
the experimental data \cite{f0-decay}. 
Radiative and two-photon decay widths
also indicate multiquark structure \cite{gamma-f0-a0}. 
Therefore, $f_0$ and $a_0$ could be considered 
as exotic tetraquark hadrons or $K\bar K$ molecules.
The $a_0$-$f_0$ mixing intensity also provides 
a clue for their $K \bar K$ compositeness \cite{sk-2014}.

Furthermore, $\Lambda (1405)$ has been considered as an exotic hadron
because the $\Lambda (1405)$ mass is anomalously light.  
Although the ground-state $\Lambda$ is heavier than the nucleon 
due to the heavier strange-quark mass, their lowest excitation states 
with $(1/2)^-$ show the reversed mass relation
$m_{\Lambda(1405)} < m_{N(1535)}$, which is difficult 
to be understood within the quark model.
It is likely to be a $\bar KN$ state.
It was pointed out in Ref. \cite{sk-2014} that a future
measurement of the $\Lambda (1405)$ radiative decay width 
should constrain the $\bar KN$ compositeness.

These hadrons have been investigated in global observables such as
spins, parities, masses, and decay widths. For finding clear evidences
of their exotic signatures, we consider to use high-energy
reactions. Because quark and gluon degrees of freedom are relevant 
at high energies, internal configurations are expected to become apparent.
Parton distribution functions are not directly measured
because unstable exotic hadrons cannot be used as fixed targets.
Then, it was studied that fragmentation functions could be 
used for such a purpose by considering the difference between
favored and disfavored functions \cite{framentaion}.
Here, we propose to use high-energy exclusive reactions 
for probing the internal structure of exotic hadron candidates.
First, we explain that the constituent counting rule can be used
for exotic hadron candidates in Sec.\,\ref{counting} by the scaling
behavior of exclusive cross sections \cite{kks-2013}.
Second, we show in Sec.\,\ref{gpd-gda}
that the generalized parton distributions (GPDs) 
and generalized distribution amplitudes (GDAs) can be used 
for tomography of exotic hadron candidates \cite{kk-2014}.

\section{Constituent-counting rule for hard exclusive production of
an exotic hadron}
\label{counting}

\begin{figure}[b]
\vspace{-0.2cm}
\begin{minipage}{0.48\textwidth}
   \begin{center}
     \includegraphics[width=3.5cm]{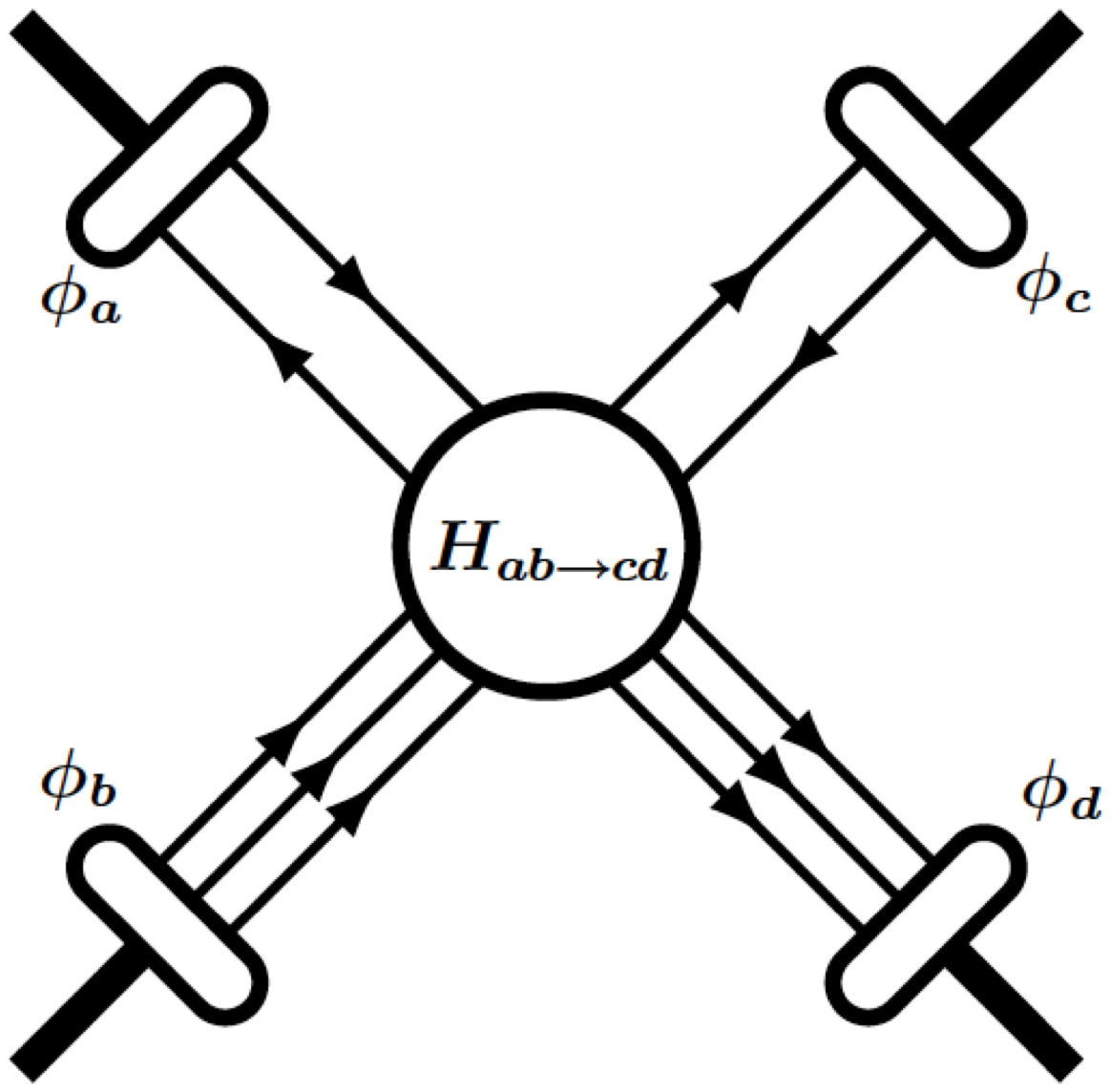}
   \end{center}
\vspace{-0.2cm}
\caption{Exclusive process $a+b \to c+d$ \cite{kks-2013}.}
\label{fig:exclusive-ab-cd}
\end{minipage}
\hspace{0.5cm}
\begin{minipage}{0.48\textwidth}
   \begin{center}
     \includegraphics[width=4.5cm]{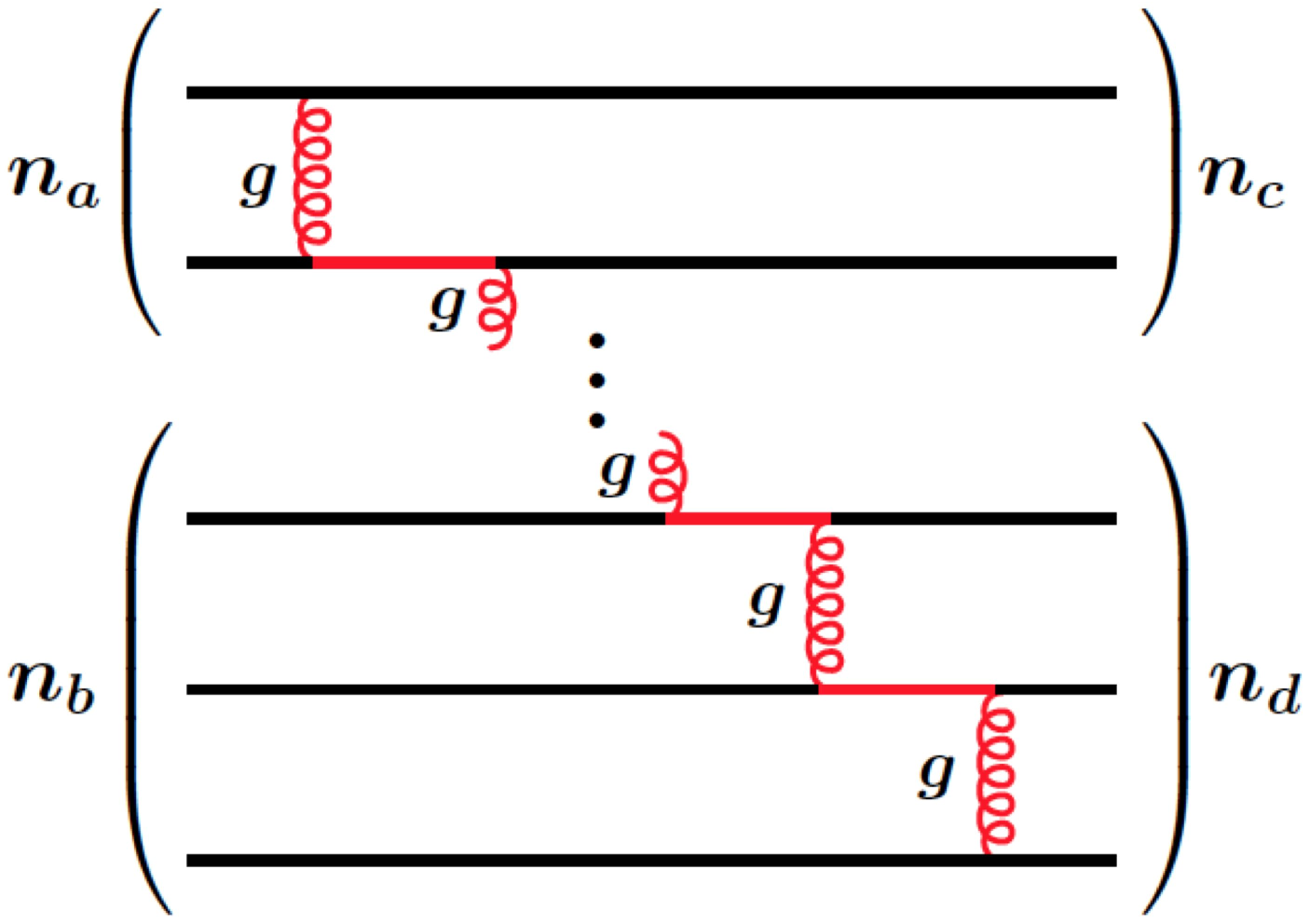}
   \end{center}
\vspace{-0.2cm}
 \caption{Hard gluon exchange process for exclusive reaction
 \cite{kks-2013}.}
\label{fig:hard-glun-exchange}
\end{minipage} 
\end{figure}

The cross section of a large-angle exclusive scattering $a+b \to c+d$
is given by
$ d\sigma_{ab \to cd} / dt \simeq
  \overline{\sum}_{pol} \, | M_{ab \to cd} |^2 / (16 \pi s^2) $,
where $s$ and $t$ are Mandelstam variables defined by
the momenta $p_h \ (h=a,\,b,\,c,\,d)$ as
$s = (p_a + p_b)^2$ and $t = (p_a - p_c)^2$.
The matrix element is expressed as
\cite{exclusive-theory}
\begin{align}
M_{ab \to cd} = & \int [dx_a] \, [dx_b] \, [dx_c] \, [dx_d]  \,
    \phi_c ([x_c]) \, \phi_d ([x_d]) 
\nonumber \\[-0.1cm]
& \ \ \ 
\times 
H_{ab \to cd} ([x_a],[x_b],[x_c],[x_d],Q^2) \, 
      \phi_a ([x_a]) \, \phi_b ([x_b]) ,
\label{eqn:mab-cd}
\end{align}
where $H_{ab \to cd}$ is the partonic scattering amplitude,
and $\phi_h$ is the light-cone distribution amplitude of 
the hadron $h$ as illustrated in Fig. \ref{fig:exclusive-ab-cd}.
A set of the light-cone momentum fractions,
$x_i=p_i^+/p_h^+$ with $i$-th parton momentum
$p_i$, is denoted $[x_h]$ for partons in a hadron $h$.

The high-energy behavior of the cross section is described by
the constituent counting rule in perturbative QCD.
As shown in Fig. \ref{fig:hard-glun-exchange},
quarks should share large momenta, by exchanging hard gluons,
so that they should stick together to form a hadron
in a large-angle-exclusive reaction.
Assigning hard momentum factors for the internal quarks,
gluons, and external quarks, we obtain the constituent-counting rule
for the cross section in perturbative QCD:
\begin{align}
\frac{d\sigma_{ab \to cd}}{dt} = \frac{1}{s^{\, n-2}} \, f_{ab \to cd}(t/s).
\label{eqn:cross-counting}
\end{align}
Here, the factor $n$ is the number of constituents defined by 
$n = n_a+n_b+n_c+n_d$, and $f(t/s)$ is a scattering-angle 
dependent function.
This expression indicates that the cross section 
is proportional to $1/s^{\, n-2}$ with the number 
of constituents. This scaling behavior 
is called the constituent-counting rule. 

\begin{figure}[h!]
\vspace{-0.4cm}
\begin{minipage}{0.48\textwidth}
   \begin{center}
     \includegraphics[width=6.0cm]{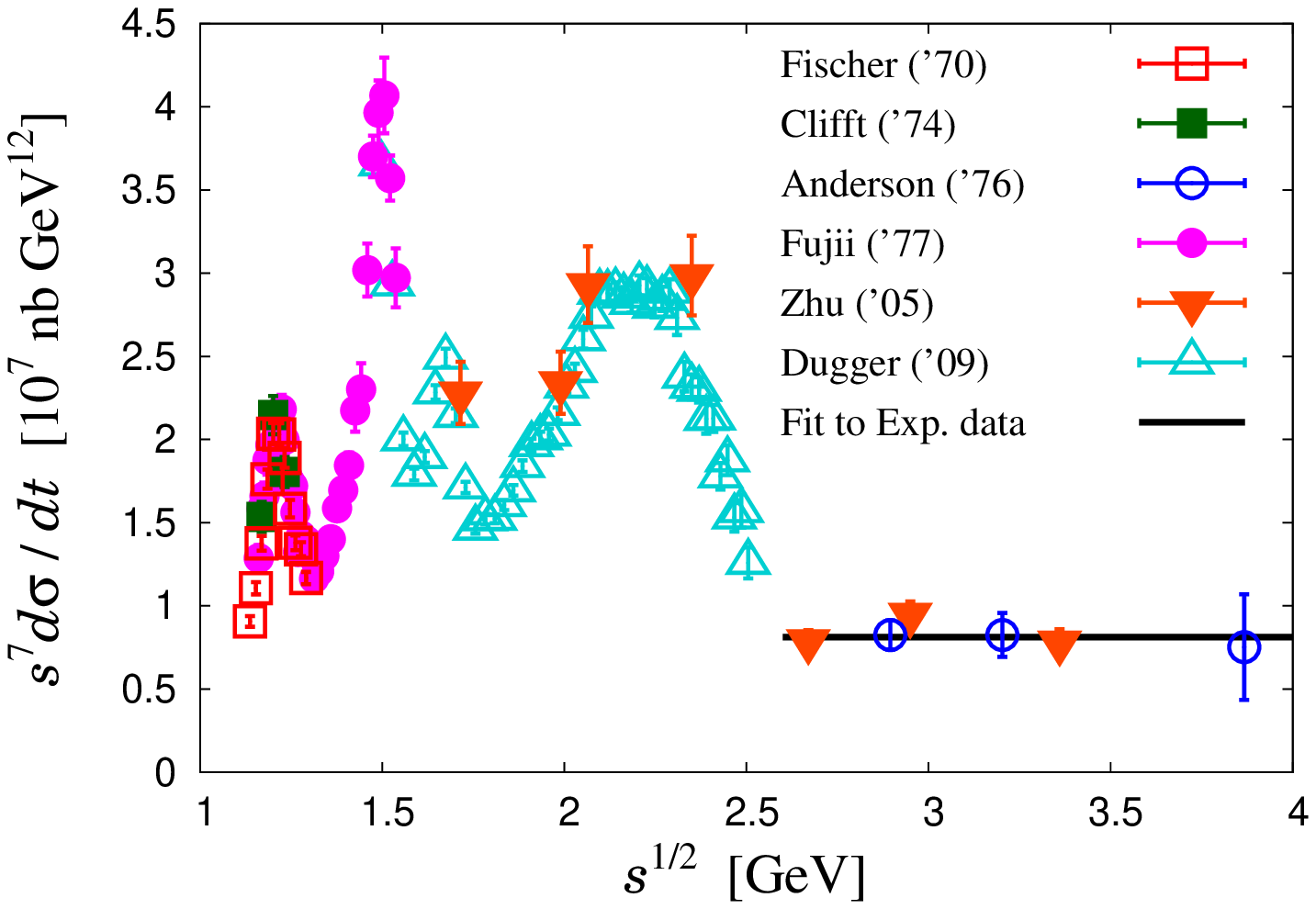}
   \end{center}
\vspace{-0.3cm}
\caption{Cross section of $\gamma + p \to \pi^+ + n$
and scaling at large energies \cite{kks-2013}.}
\label{fig:gamma-p}
\end{minipage}
\hspace{0.5cm}
\begin{minipage}{0.48\textwidth}
   \begin{center}
     \includegraphics[width=5.85cm]{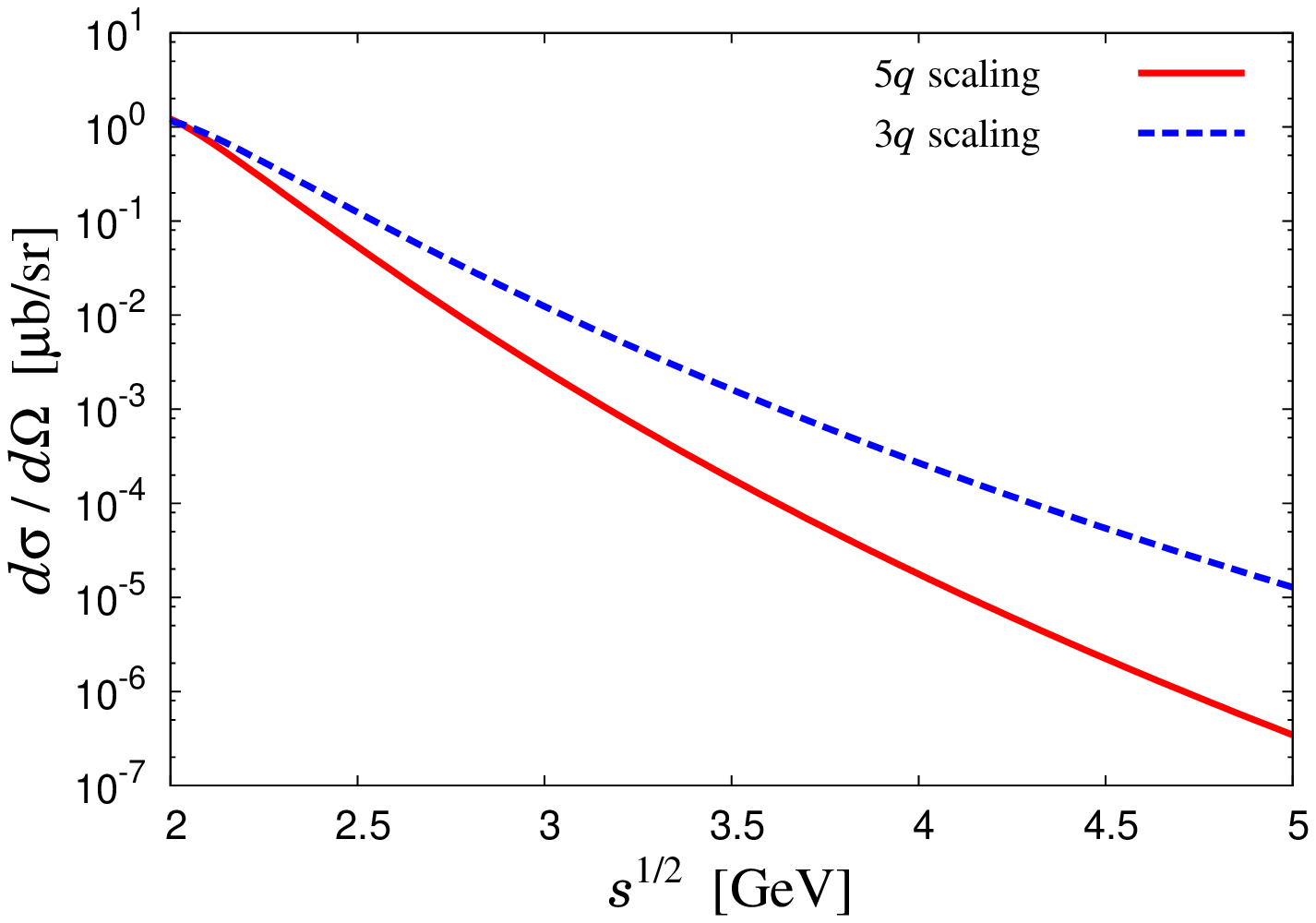}
   \end{center}
\vspace{-0.3cm}
\caption{Scaling of
         $\pi^- + p \to K^0 + \Lambda \, (1405)$ cross section
         at high energies \cite{kks-2013}.}
\label{fig:lambda-1405-scaling}
\end{minipage} 
\end{figure}

The counting rule has been experimentally investigated
\cite{counting-exp}.
In various hadron two-body reactions, BNL measurements
support the scaling predicted by the counting rule.
Furthermore, it is also indicated in lepton-facility
measurements. In Fig. \ref{fig:gamma-p}, 
the cross section of $\gamma + p \to \pi^+ + n$ is shown
at $\theta_{cm}=90^\circ$ as a function of $\sqrt{s}$.
Here, the cross section is multiplied by the factor
$s^7$ because the number is $n-2=7$. 
The figure shows typical resonance structure at low energies,
whereas the cross section is almost constant at high energies.
The data indicate that the transition from hadron degrees
of freedom to the quark and gluon ones occurs at 
$\sqrt{s}=2.5$ GeV. 

The scaling of the cross section could be used for
probing internal structure of an exotic hadron because
the slope is controlled by the number of constituents ($n$)
\cite{kks-2013}.
As an example, we estimate the cross section of $\Lambda (1405)$ 
production, $\pi^- + p \to K^0 +\Lambda (1405)$.
For this purpose, we first investigate the ground-$\Lambda$ production
$\pi^- + p \to K^0 +\Lambda$, for which there are many available data
at $\theta_{cm}=90^\circ$. The cross section data of 
$\pi^- + p \to K^0 +\Lambda$ show the scaling with
$n=10.1 \pm 0.6$ at high energies, and it is consistent
with the number of constituents $n=2+3+2+3=10$.
Therefore, it is promising to use the counting rule
also for $\Lambda (1405)$ for finding whether it
is an ordinary $qqq$ state of five-quark one.
There is only one experimental data available for 
the $\Lambda (1405)$ cross section at $\theta_{cm}=90^\circ$.
We use this data together with the counting rule
to show the scaling of the cross section at high energies
in Fig. \ref{fig:lambda-1405-scaling} 
by assuming that the perturbative region starts at $\sqrt{s}=2$ GeV. 
Two curves are shown by assuming
three- or five-quark state for $\Lambda (1405)$.
There are large differences between the two curves as
the energy becomes larger, which should be used for judging
whether $\Lambda (1405)$ is an exotic five-quark state.
Such an experiment is possible, for example, at J-PARC.

\vspace{-0.8cm}


\section{Tomography of exotic hadrons by generalized parton distributions
and generalized distribution amplitudes}
\label{gpd-gda}

\begin{wrapfigure}[9]{r}{0.45\textwidth}
   \vspace{-0.80cm}
   \begin{center}
     \includegraphics[width=5.5cm]{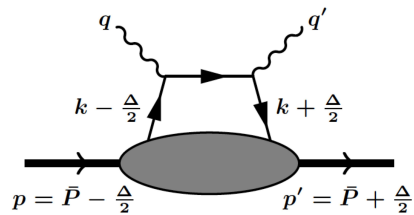}
   \end{center}
\vspace{-0.1cm}
\caption{Virtual Compton scattering for GPD \cite{kk-2014}.}
\label{fig:gpd-1}
\vspace{-0.5cm}
\end{wrapfigure}

We proposed that the internal quark configuration could be
found by looking at the scaling of an exclusive cross section
at high energies. For finding another independent evidence
at high energies and for probing much details of internal
structure in the three-dimensional form, we propose to
use hadron tomography by using generalized parton distributions
(GPDs) and generalized distribution amplitudes (GDAs) \cite{kk-2014}.
Recently, three dimensional structure of the nucleon has been
investigated by using the GPDs and TMDs 
(transverse-momentum-dependent parton distributions)
\cite{gpd-gda-summary},
particularly for understanding the origin of nucleon spin
due to orbital angular momenta of partons.

The GPDs are measured by the virtual Compton process in Fig. \ref{fig:gpd-1}.
We define kinematical variable for expressing the GPDs.
First, the momenta $\bar P$, $\bar q$, and $\Delta$ are defined by
the nucleon and photon momenta as
$\bar P = (p+p')/2$,
$\bar q = (q+q')/2$, 
$\Delta = p'-p = q-q'$.
Then, the momentum-transfer-squared quantities are given by
$Q^2 = -q^2$, $\bar Q^2 = - \bar q^2$, and $t = \Delta^2$.
The generalized scaling variable $x$ and a skewdness parameter $\xi$ 
are defined by
$x   = Q^2/(2p \cdot q)$ and
$\xi = \bar Q^2 /(2 \bar P \cdot \bar q)$.
The variable $x$ indicates the lightcone momentum fraction 
carried by a quark in the nucleon.
The skewdness parameter $\xi$ or the momentum $\Delta$ indicates
the momentum transfer from the initial nucleon to the final one
or the one between the quarks. 

\begin{figure}[b]
\vspace{-0.2cm}
\begin{minipage}{0.48\textwidth}
   \begin{center}
     \includegraphics[width=5.0cm]{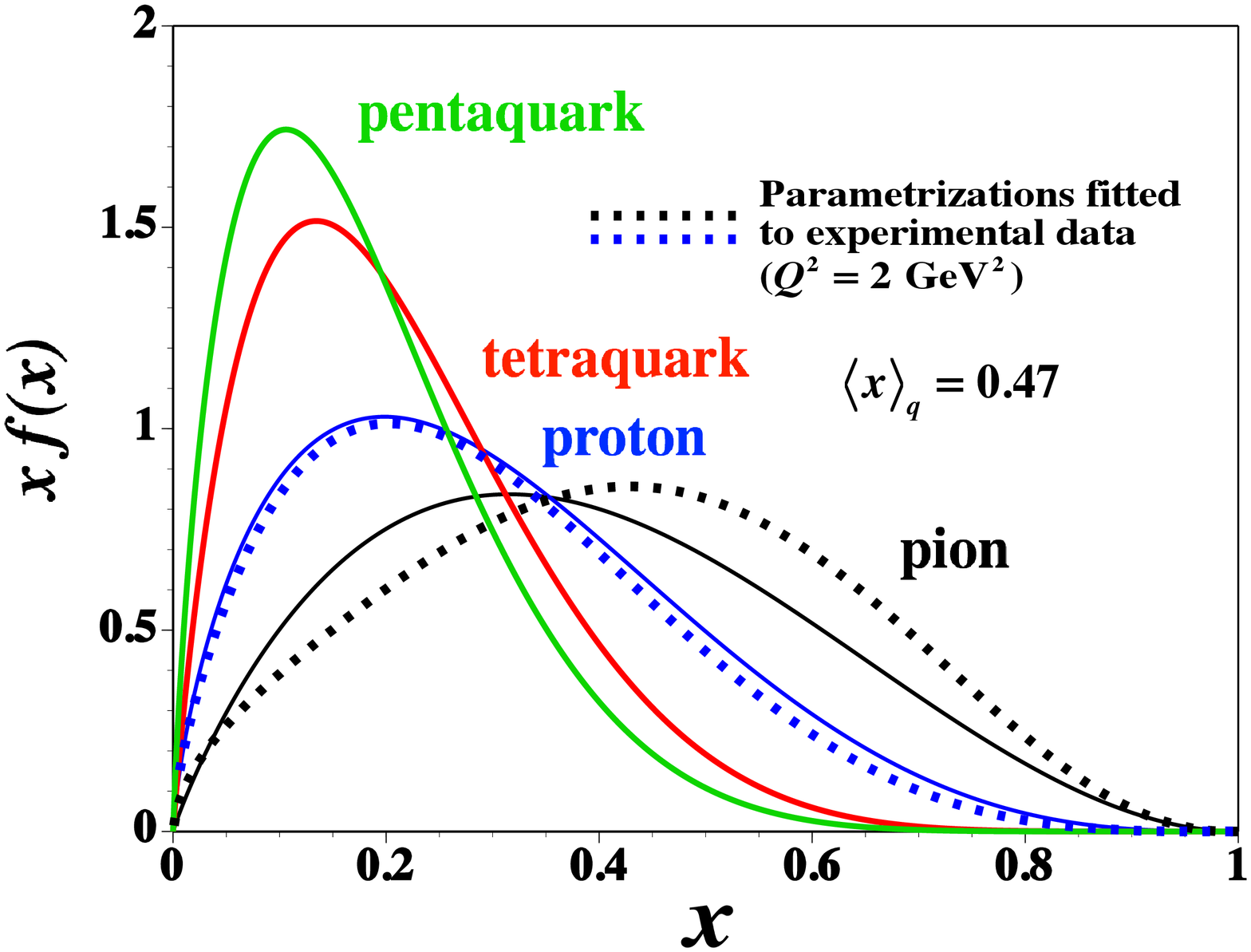}
   \end{center}
\vspace{-0.1cm}
\caption{PDFs of exotic hadrons in comparison
with parametrizations of pion and proton PDFs.
The solid curves are calculated by using a simple
function suggested by the constituent counting rule \cite{kk-2014}.
}
\label{fig:exotic-pdfs}
\end{minipage}
\hspace{0.5cm}
\begin{minipage}{0.48\textwidth}
   \begin{center}
     \includegraphics[width=5.0cm]{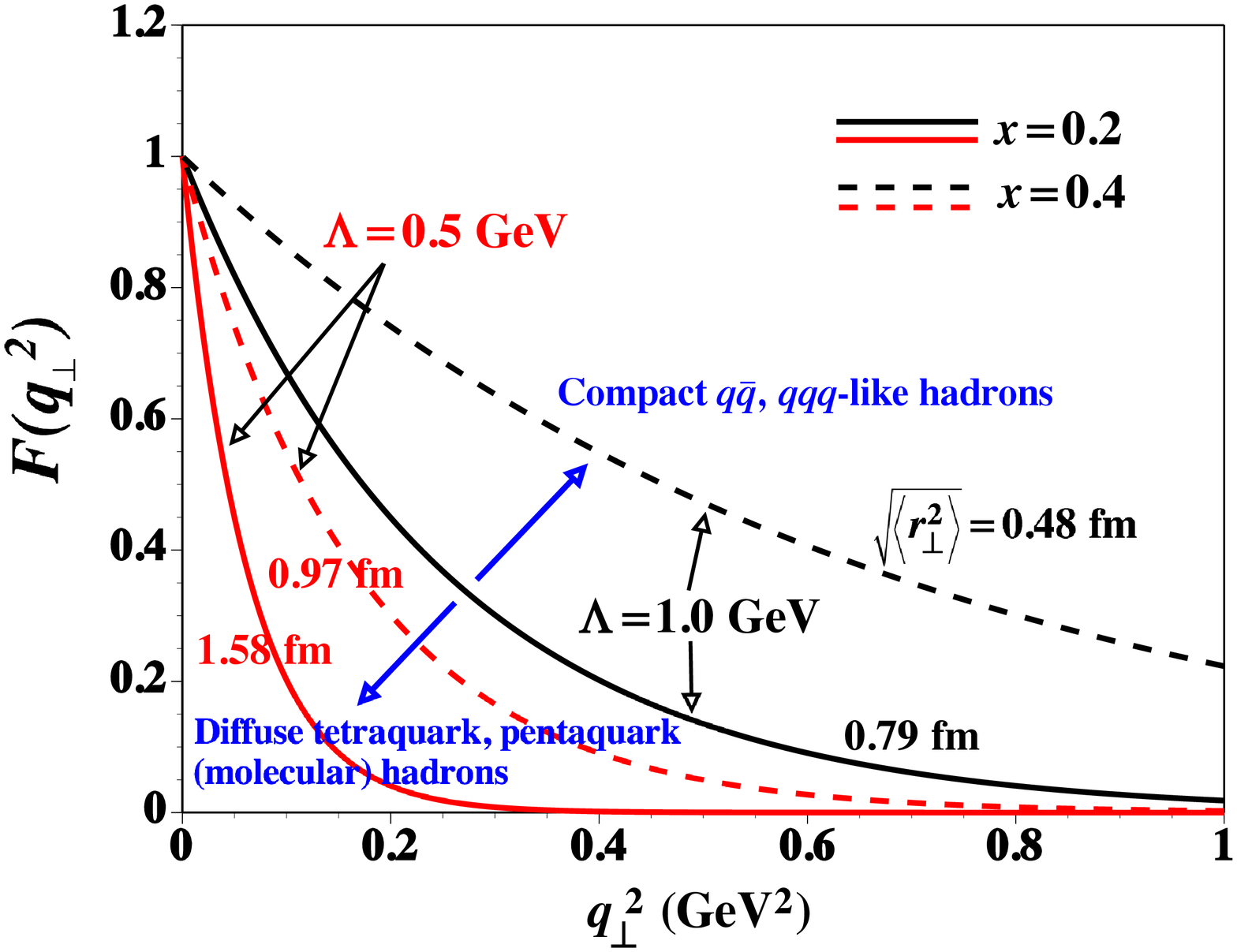}
   \end{center}
\vspace{-0.0cm}
\caption{Transverse form factors for $x=0.2$, 0.4 and
$\Lambda=0.5$, 1.0 GeV \cite{kk-2014}.}
\label{fig:transverse-form}
\end{minipage} 
\end{figure}

The GPDs for the nucleon are defined by off-forward matrix elements
of quark (and gluon) operators with a lightcone separation 
between nucleonic states:
\begin{align}
 & \! \! \! \! \! \! \! \! 
 \int\frac{d y^-}{4\pi}e^{i x \bar P^+ y^-}
 \left< p' \left| 
 \overline{\psi}(-y/2) \gamma^+ \psi(y/2) 
 \right| p \right> \Big |_{y^+ = \vec y_\perp =0}
\nonumber \\
 & \! \! \! \! \! \! \! \! \! \! \! \!
 = \frac{1}{2  \bar P^+} \, \overline{u} (p') 
 \left [ H_q (x,\xi,t) \gamma^+
     + E_q (x,\xi,t)  \frac{i \sigma^{+ \alpha} \Delta_\alpha}{2 \, M}
 \right ] u (p) .
\label{eqn:gpd-n}
\end{align}
Here, $\sigma^{\alpha\beta}$ is 
$\sigma^{\alpha\beta}=(i/2)[\gamma^\alpha, \gamma^\beta]$,
and the unpolarized GPDs of the nucleon are 
$H_q (x,\xi,t)$ and $E_q (x,\xi,t)$.
The quark field is denoted as $\psi(y/2)$, and 
$u (p)$ is the Dirac spinor.
The GPDs contain information on both longitudinal
momentum distributions of partons and transverse
form factors. It can be seen first by taking
the forward limit ($\Delta,\, \xi,\, t \rightarrow 0$):
$H_q (x, 0, 0) = q(x)$,
where $q(x)$ is an unpolarized parton distribution function (PDF)
in the nucleon.
Next, their first moments are the form factors of the nucleon:
$ \int_{-1}^{1} dx H_q(x,\xi,t)  = F_1 (t)$ and
$ \int_{-1}^{1} dx E_q(x,\xi,t)  = F_2 (t)$.
Because of these two ingredients, a useful functional form of
the GPDs 
\begin{align}
H_q^h (x,\xi=0,t)= f_n (x) \, F_n^h (t, x) ,
\label{eqn:gpd-paramet1}
\end{align}
is often used. Here, $f_n (x)$ is a longitudinal PDF,
and $F_n^h (t, x)$ is a transverse form factor at $x$.

Because there is no stable exotic hadron, the GPDs cannot be
directly measured except for transition GPDs, for example,
for $p \to \Lambda (1405)$. However, we consider 
a {\it gedankenexperiment} by which the PDFs or the GPDs 
of the exotic hadrons can be obtained
by assuming a stable target.
A simple form of the PDFs is given by
$ f_n (x) = C_n \, x^{\alpha_n} \, (1-x)^{\beta_n} $,
where the parameters are constrained by
the valence-quark number
$ \int_0^1 dx \, f_n (x) = n$
and the quark momentum
$ \int_0^1 dx \, x \, f_n (x) = \langle x \rangle_q$.
The parameter $\beta_n$ could be determined by
the constituent counting rule 
as $\beta_n = 2n -3+2\Delta S_z$ with
the spin factor $\Delta S_z=|S_z^q-S_z^h|$
because the factor
$(1-x)^{\beta_n}$ controls the behavior in the elastic
limit $x \to 1$.
Using this $\beta_n$, the number of constituents $n$,
and the momentum fraction $\langle x \rangle_q$,
we obtain $C_n$ and $\alpha_n$.
The calculated PDFs are shown in Fig. \ref{fig:exotic-pdfs}
for $n=2$ (meson), 3 (baryon), 4 (tetraquark), and 5 (pentaquark).
In comparison, typical PDFs determined from experimental data
are shown by the dashed curves for the pion and nucleon 
at $Q^2=2$ GeV$^2$.
Although differences between the dashed and solid curves 
vary depending on the $Q^2$ value, we obtain a reasonable
agreement. It means that the PDFs obtained by the counting rule
are reasonable magnitude estimates.
As the valence-quark number becomes larger ($n=4$, 5),
the distribution shifts toward the small-$x$ region.
In Fig. \ref{fig:transverse-form}, transverse form factors
are shown by assuming the exponential form
$ F_n^h (t,x) = e^{(1-x) t/(x \Lambda^2)} $
with the cutoff parameter constrained by the transverse size as
$ \langle  r_\perp^2 \rangle = 4(1-x) / (x \Lambda^2) $.
The $q_\perp^2$ dependence changes significantly whether
the hadron has compact $q\bar q$ ($qqq$)-like structure 
or diffuse tetraquark (pentaquark, hadron molecule) one.

\begin{wrapfigure}[12]{r}{0.43\textwidth}
   \vspace{-0.9cm}
   \begin{center}
     \includegraphics[width=5.5cm]{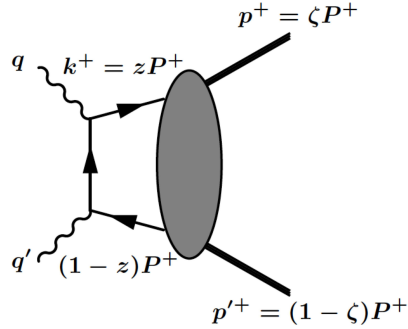}
   \end{center}
\vspace{-0.3cm}
\caption{$\gamma^* \gamma \to h \bar h$ process for GDA \cite{kk-2014}.}
\label{fig:gda-1}
\vspace{-0.5cm}
\end{wrapfigure}

The GDAs are defined in the same way with the GPDs 
in the $s$-$t$ crossed channel as shown in Fig. \ref{fig:gda-1}.
They describe the production of a hadron pair $h\bar h$,
$\gamma^* \gamma \to h \bar h$. The final-hadron momenta are
denoted $p$ and $p'$, the initial photon momenta are 
$q$ ($Q^2=-q^2$) and $q'$  (${q'}^2=0$). 
$P$ is the total momentum $P=p+p'$, and $k$ is the quark momentum. 
The center-of-mass (c.m.) energy squared $s$ is equal to
the invariant-mass squared $W^2$ of the final hadron pair,
$s= (q+q')^2 = (p+p')^2 = W^2$.
Then, the variable $\zeta$ is defined by
$ \zeta = p \cdot q' / (P \cdot q')
        = p^+ / P^+ = (1+\beta \cos\theta)/2 $,
where $\beta$ is the velocity of a hadron,
$ \beta = |\vec p \, | / p^0 = ( 1-4 m_h^2 / W^2 )^{1/2} $,
with the final hadron mass $m_h$, and
$\theta$ is the scattering angle in the c.m. frame.
The GDAs are expressed
by these three variables ($z$, $\zeta$, $s=W^2$).

The GDAs are defined by the same lightcone operators
between the vacuum and the hadron pair $h\bar h$:
\begin{align}
\Phi_q^{h\bar h} (z,\zeta,s) 
= \int \frac{d y^-}{2\pi}\, e^{i (2z-1)\, P^+ y^-}
   \langle \, h(p) \, \bar h(p') \, | \, 
 \overline{\psi}(-y/2) \gamma^+ \psi(y/2) 
  \, | \, 0 \rangle \Big |_{y^+=\vec y_\perp =0} \, ,
\end{align}
for a quark. The gluon GDA is defined in the similar way.
The GDAs are related to the GPDs by the $s$-$t$ crossing
if the factorizations can be applied as shown
in Figs. \ref{fig:gpd-1} and \ref{fig:gda-1}.
The crossing means to move the final state $\bar h$ ($p'$) 
to the initial $h$ ($p$). It indicates that the momenta ($p$, $p'$)
of the GDAs should be replaced by ($p'$, $-p$) in the GPDs.
Then, the relations between the variables are given by
$ z \leftrightarrow (1-x/\xi)/2$,  
$ \zeta \leftrightarrow (1-1/\xi)/2$, and
$ W^2 \leftrightarrow t$,
which indicates that the GDAs are related to the GPDs by
\begin{align}
\Phi_q^{h\bar h} (z,\zeta,W^2) 
\longleftrightarrow
H_q^h \left ( x=\frac{1-2z}{1-2\zeta},
            \xi=\frac{1}{1-2\zeta}, t=W^2 \right ) .
\label{eqn:gda-gpd-relation}
\end{align}
Because there are many studies on the GPDs, this relation seems to 
be useful for estimating the GDAs. However, we find that the relevant
kinematical regions are 
$ 0 \le |x| < \infty$, $ 0 \le |\xi| < \infty$, 
$ |x| \le |\xi|$,      $ t \ge 0 $, 
which are not necessarily the usual physical regions of the GDAs,
so that the information of the GPDs is not used directly
for the GDAs.

The cross section for $e\gamma  \to e h \bar h$ is given by
\cite{diehl-2000}
\begin{align}
\frac{{d\sigma }}{{d{Q^2}d{W^2}d\cos \theta}} 
= \frac{\beta \, \alpha ^3}{8 s_{e\gamma}^2 \, Q^2 (1 - \varepsilon ) }
{\left| {{A_{+ +}}(\zeta ,{W^2})} \right|^2} , \ \ 
A_{++} = \sum\limits_q {\frac{{e_q^2}}{2}} \int_0^1 {dz} 
\frac{{2z - 1}}{{z(1 - z)}}\Phi _q^{h \bar h}(z,\zeta ,{W^2}) ,
\label{eqn:corss-2}
\end{align}
in the leading order of $\alpha_s$ by neglecting gluon GDAs.
Here, $A_{+ +}$ is the helicity amplitude 
$ A_{i  j}= \varepsilon _\mu ^{( i )}(q) \, 
  \varepsilon _\nu ^{( j )}(q') \, {T^{\mu \nu }} /e^2 $
for the hadron tensor $T^{\mu \nu }$ of $\gamma^* \gamma \to h \bar h$, and
$\Phi _q^{h \bar h}$ is a quark GDA.
In oder to find possible signatures of exotic hadrons, we
use a simple function
\begin{align}
\Phi_q^{h\bar h (I=0)} (z,\zeta,W^2) 
= N_{h(q)} \, z^\alpha (1-z)^\beta (2z-1) \, \zeta (1-\zeta) \, F_{h(q)} (W^2) ,
\label{eqn:gda-paramet}
\end{align}
which satisfies the sum rules for the quark GDAs for 
the isospin $I=0$ two-meson final states:
\begin{align}
\int_0^1 dz \,  \Phi_q^{h\bar h (I=0)} (z,\zeta,W^2) = 0, \ \ 
\int_0^1 dz \,  (2z -1) \, \Phi_q^{h\bar h (I=0)} (z,\zeta,W^2) 
= - 2 M_{2(q)}^h \zeta (1-\zeta) F_{h(q)} (W^2) ,
\label{eqn:gda-sum-I=0}
\end{align}
\begin{wrapfigure}[12]{r}{0.42\textwidth}
   \vspace{-0.80cm}
   \begin{center}
     \includegraphics[width=5.6cm]{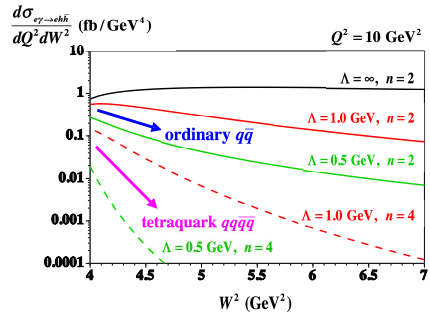}
   \end{center}
\vspace{-0.35cm}
\caption{Form-factor effects on the cross section 
as a function of $W^2$ at $Q^2=$10 GeV \cite{kk-2014}.}
\label{fig:cross-form}
\vspace{-0.5cm}
\end{wrapfigure}
\noindent
where $M_{2(q)}^h$ is the momentum fraction carried by quarks.
The function $F_{h(q)} (W^2)$ is a form factor of the quark part
of the energy-momentum tensor, and it may be taken in the form
suggested by the counting rule: 
$ F_{h(q)} (W^2) 
= 1 / [ 1 + (W^2-4 m_h^2)/\Lambda^2 ]^{n-1} $,
where $n=2$ for ordinary $q\bar q$ mesons and $n=4$ 
for tetraquark hadrons.
Using the form factor together with the GDA expression in
Eq. (\ref{eqn:gda-paramet}), we obtain the cross section
for $e+\gamma \to e + h + \bar h$ as the function of $W^2$
in Fig. \ref{fig:cross-form} by considering $h=f_0 (980)$ or $a_0 (980)$
\cite{kk-2014}.
Depending on the size and the constituent number, the cross section
varies much as the function of $W^2$. Therefore, 
measurements of the $e\gamma \to e' h\bar h$ cross section
should be valuable for determining internal structure of
exotic hadrons through the GDAs.


\section{Summary}
\vspace{-0.2cm}
Exotic hadron candidates have been investigated in hadron spectroscopy and their 
decays; however, the high-energy processes can probe their internal structure.
In this work, we studied exclusive reactions by using the idea of
constituent counting rule and by the hadron tomography 
with the GPDs and GDAs. Because the quark and gluon degrees of freedom
are relevant at high energies, it should be more appropriate to use 
the high-energy exclusive reactions for probing the internal configurations
of exotic hadron candidates.


\vspace{-0.0cm}
\section*{Acknowledgements}
\vspace{-0.2cm}
This work was supported by the MEXT KAKENHI Grant Number 25105010.




\begin{thebibliography}{9}
\bibitem{pdg-2014} K. A. Olive {\it et al}. 
                    Chin. Phys. C {\bf 38}, 090001 (2014).
\bibitem{f0-decay} S. Kumano and V. R. Pandharipande, 
                              Phys. Rev. D {\bf 38}, 146 (1988).
\bibitem{gamma-f0-a0} F. E. Close, N. Isgur, and S. Kumano, 
                              Nucl. Phys. B {\bf 389}, 513 (1993);
            N. N. Achasov and G. N. Shestakov,
                  Nucl. Phys. Proc. Suppl. {\bf 225-227}, 135 (2012).
\bibitem{sk-2014} T. Sekihara and S. Kumano, 
                        Phys. Rev. C {\bf 89}, 025202 (2014); 
                        arXiv:1409.2213.
\bibitem{framentaion} M. Hirai, S. Kumano, M. Oka, and K. Sudoh,
                              Phys. Rev. D {\bf 77}, 017504 (2008). 
\bibitem{kks-2013} H. Kawamura, S. Kumano, and T. Sekihara, 
                        Phys. Rev. D {\bf 88}, 034010 (2013).
\bibitem{kk-2014} H. Kawamura and S. Kumano,
                        Phys. Rev. D {\bf 89}, 054007 (2014).
\bibitem{exclusive-theory}
    S. J. Brodsky and G. P. Lepage, Adv.\ Ser.\ Direct.\ 
                  High Energy Phys.\  {\bf 5}, 93 (1989).
\bibitem{counting-exp} 
   C. White {\it et al.}, Phys. Rev. D {\bf 49}, 58 (1994);
   L. Y. Zhu {\it et al.}, Phys. Rev. C {\bf 71}, 044603 (2005).
\bibitem{gpd-gda-summary} 
            M. Diehl, Phys. Rept. {\bf 388}, 41 (2003);
            M. Diehl and P. Kroll, Eur. Phys. J. C {\bf 73}, 2397 (2013).
\bibitem{diehl-2000} M. Diehl, T. Gousset, and B. Pire, 
                        Phys. Rev. Lett. {\bf 81}, 1782 (1998);
                        Phys. Rev. D {\bf 62}, 073014 (2000).
\end{thebibliography}
\end{document}